\documentclass[12pt]{article}

\usepackage{epsfig}

\begin{document}

\title{\Large \bf Information theory point of view on multiparticle
                  production processes }
\author{\large O.V.Utyuzh$^1$\thanks{e-mail: utyuzh@fuw.edu.pl},
               G.Wilk$^1$\thanks{e-mail: wilk@fuw.edu.pl},
               Z.W\l odarczyk$^2$\thanks{e-mail: wlod@pu.kielce.pl}
                \bigskip \\
{\it  $^1$~The Andrzej So\l tan Institute for Nuclear Studies,}\\{\it
Ho\.za 69, 00681 Warsaw, Poland} \\  {\it $^2$~Institute of Physics,
\'Swi\c{e}tokrzyska Academy,}\\{\it \'Swi\c{e}tokrzyska 15; 25-406
Kielce, Poland}}

\maketitle


\begin{center}
{\bf Abstract}
\end{center}
\medskip
We review hitherto attempts to look at the multiparticle production
processes from the Information Theory point of view  (both in its
extensive and nonextensive versions). \\




\section{Why Information Theory?} \label{s1}

The idea of using Information Theory (IT) to analyze  the multiparticle
production data is very old one and can be traced back to \cite{Chao}.
This work attempted to establish what the experimental data of that time
are telling us. The problem was serious because there was a number of
theoretical models based on apparently completely disparate physical
assumptions, all of which were claiming to provide fairly good
description of data and therefore were in fierce competition between
themselves. The working hypothesis of \cite{Chao} was that experimental
data contain only limited amount of {\it information}, which was {\it
common} to all of them. Models considered differed therefore in some
other aspects, which, from the point of experimental data considered,
were however {\it irrelevant}. As it turned out, this information was
that: $(i)$ not all available energy $\sqrt{s}$ is used for production of
secondaries (i.e., existence of {\it inelasticity}) and $(ii)$ that
transverse momenta od produced particles are strongly damped (apparent
one-dimensionality of the relevant phase space). After closer scrutiny of
models it was revealed that, indeed, all of them have these features
build in (some in explicit some in very implicit way). From the point of
view of data all these models were simply equivalent and their
differences in what concerns their other physical assumptions were simply
nonexistent.

\section{Basics of IT approach}

This result could be obtained only by quantifying the notion of
information, i.e., by resorting to IT with all its mathematical
machinery. Referring to \cite{MaxEnt} for list of the relevant references
we shall briefly sketch its basics. To do this one has first to introduce
some {\it measure of information} contained in a given probability
distribution $p_i$. It is usually provided Shannon information entropy
defined as:
\begin{equation}
S\, =\, -\Sigma_i p_i\ln p_i .\label{eq:Shannon}
\end{equation}
Actually the choice of entropy used as measure of information reflects
our {\it a priori} knowledge about hadronizing system. Out of many
possibilities \cite{ENT} we shall present here results obtained using the
so called Tsallis entropy \cite{T},
\begin{equation}
S_q\, =\, - \frac{1}{1-q}\Sigma_i\left(1 - p_i^q\right)
.\label{eq:Tsallis}
\end{equation}
It is characterized by the so called nonextensivity parameter $q$, which
quantifies deviation from the usual additivity of the entropy, namely
that for $S_q$, for two independent systems $A$ and $B$
\begin{equation}
S_{q(A+B)} = S_{qA} + S_{qB} + (1-q)S_{qA}S_{qB}. \label{eq:SAB}
\end{equation}
Notice that in the limit $q\rightarrow 1$ one recovers the previous form
of Boltzmann-Gibbs-Shannon entropy (\ref{eq:Shannon}).

The IT tells us that the available data should be described by using only
a {\it truly minimal amount of information} in order to avoid any
unfounded and unnecessary assumptions. This information is provided by a
finite number of observables $\{ F_k, k=1,\dots, n\}$ of some physical
quantities obtained by means of $p_i$ and defined as:
\begin{equation}
\langle {F}_k\rangle = \sum_{i=1}^{n} p_i\, {F}_k^{(i)}
,\label{eq:average}
\end{equation}
when Shannon entropy is used and as:
\begin{equation}
\langle {F}_k\rangle_q = \sum_{i=1}^{n} [p_i]^q\, {F}^{(i)}_k
,\label{eq:qaverage}
\end{equation}
when one uses Tsallis entropy instead. With this information one is then
looking for (normalized) probability distribution $\{ p_i\}$ which
contains only information provided by $\{F_k\}$ and nothing more, i.e.,
which contain {\it minimal} information. As minimal information means
maximal corresponding information entropy one is therefore looking for
$\{p_i\}$ maximizing this entropy subjected to the to constraints given
by eqs.(\ref{eq:average}) or (\ref{eq:qaverage}) which account for our
{\it a priori} knowledge of the process under consideration. As a result
one gets the {\it most probable} and {\it least biased} distribution
describing these data, which is not influenced by anything else besides
the available information. In case of Shannon information entropy it
is\footnote{Notice that using the entropic measure $S\, =\,
\sum_i\left[p_i \ln p_i\, \mp\, \left(1\, \pm\, p_i\right) \ln \left(1\,
\pm\, p_i \right) \right]$ (which, however, has nothing to do with {\it
IT}) would result instead in Bose-Einstein and Fermi-Dirac formulas:
$p_i\, =\, (1/Z)\cdot \left[\exp[\beta(\varepsilon_i - \mu)] \mp
1\right]^{-1}$, where $\beta$ and $\mu$ are obtained solving two
constraint equations given, respectively, by energy and number of
particles conservation \cite{TMT}. It must be also stressed that the
final functional form of $p_i$ depends also on the functional form of the
constraint functions $F_k (x_i)$. For example, $F(x) \propto \ln (x)$ and
$\ln (1-x)$ type constrains lead to $p_i \propto
x_i^{\alpha}(1-x_i)^\beta$ distributions.}
\begin{equation}
p_i = \frac{1}{Z}\, \exp\left[ - \sum_{k=1}^r\, \lambda_k \cdot F_k^{(i)}
\right], \label{eq:pi}
\end{equation}
(where $Z$ is obtained from the normalization condition $\sum_{i=1}^np_i
= 1$) whereas in case of Tsallis entropy it is
\begin{equation}
p_i\, =\, p^{(q)}_{i}\, =\, \frac{1}{Z_q}\, \exp_q\left[ - \sum_{k=1}^r\,
\lambda_k \cdot F^{(i)}_k \right], \label{eq:qpi}
\end{equation}
where $Z_q$ is obtained from the normalization condition
$\sum_{i=1}^np^{(q)}_i = 1$ and where
\begin{equation}
\exp_q \left(- \frac{x}{\Lambda}\right) \stackrel{\it def}{=} \left[1 -
(1-q) \left(\frac{x}{\Lambda}\right)\right]^{\frac{1}{1-q}}
.\label{eq:qexp}
\end{equation}

\section{Examples of applications}

\begin{figure}[h]
\begin{center}
  \begin{minipage}[ht]{5.3cm}
    \centerline{\epsfig{file=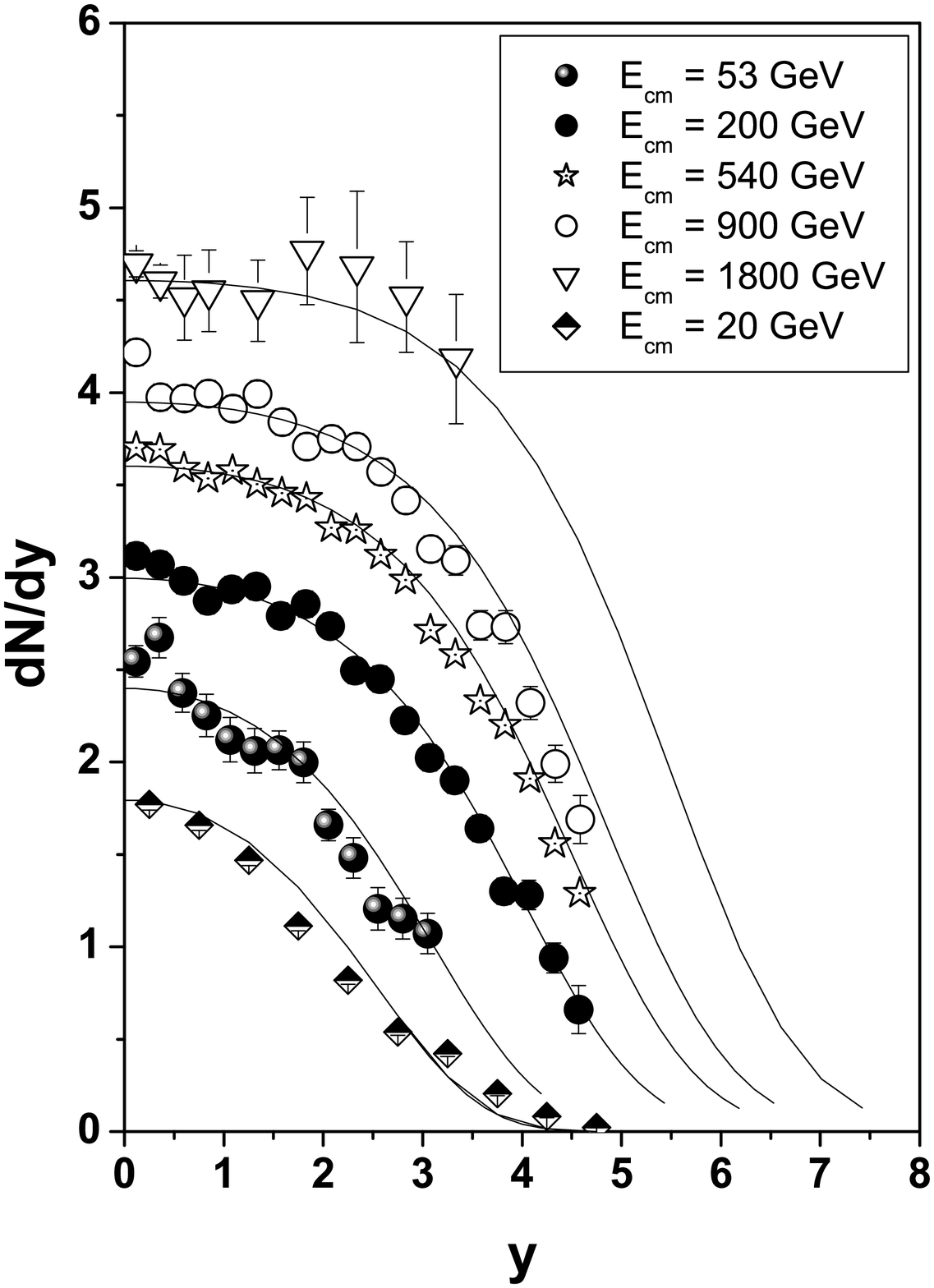, width=53mm}}
  \end{minipage}
\hfill
  \begin{minipage}[ht]{5.3cm}
    \centerline{\epsfig{file=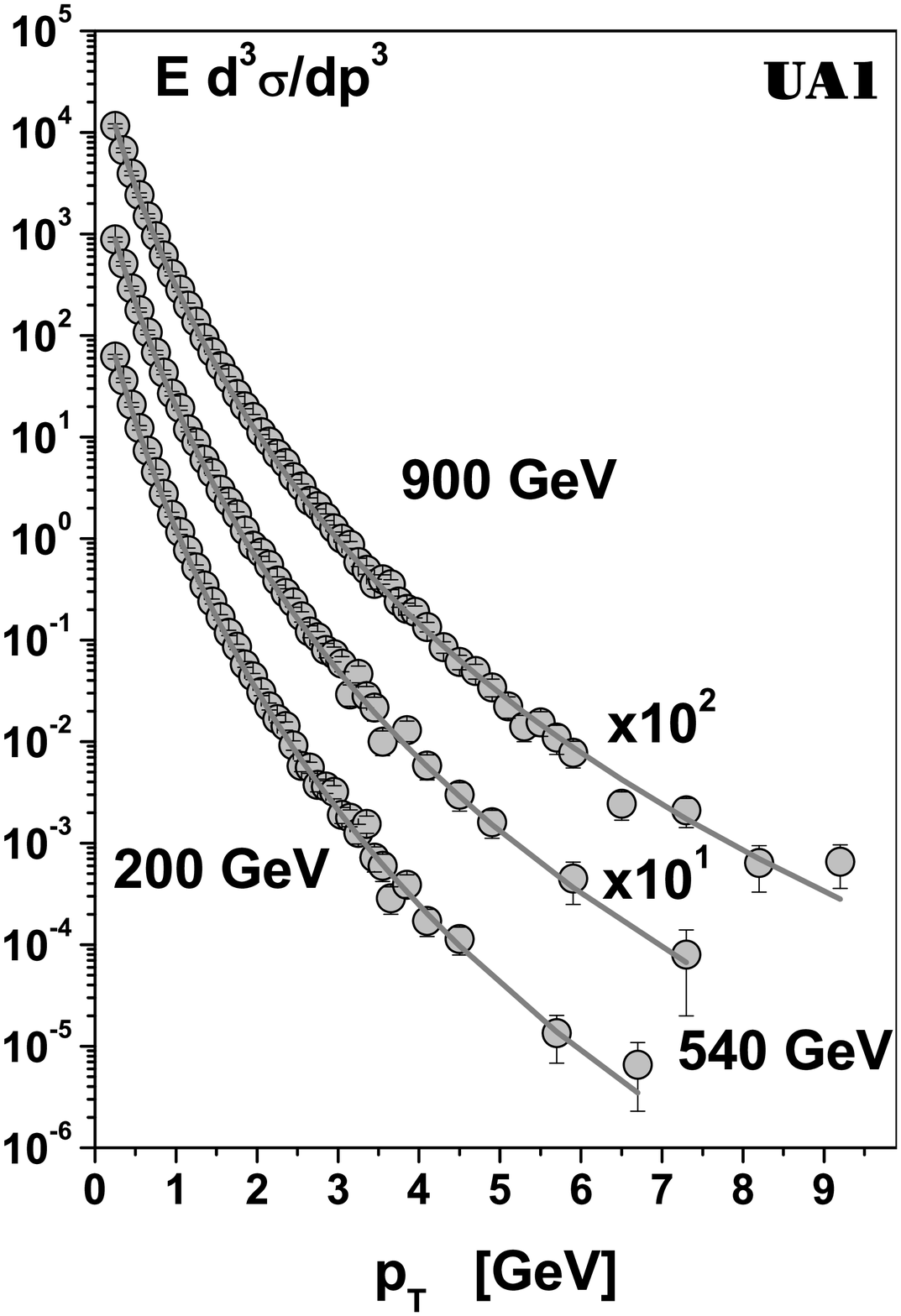, width=51mm}}
  \end{minipage}
\hfill
  \begin{minipage}[ht]{5.3cm}
    \centerline{\epsfig{file=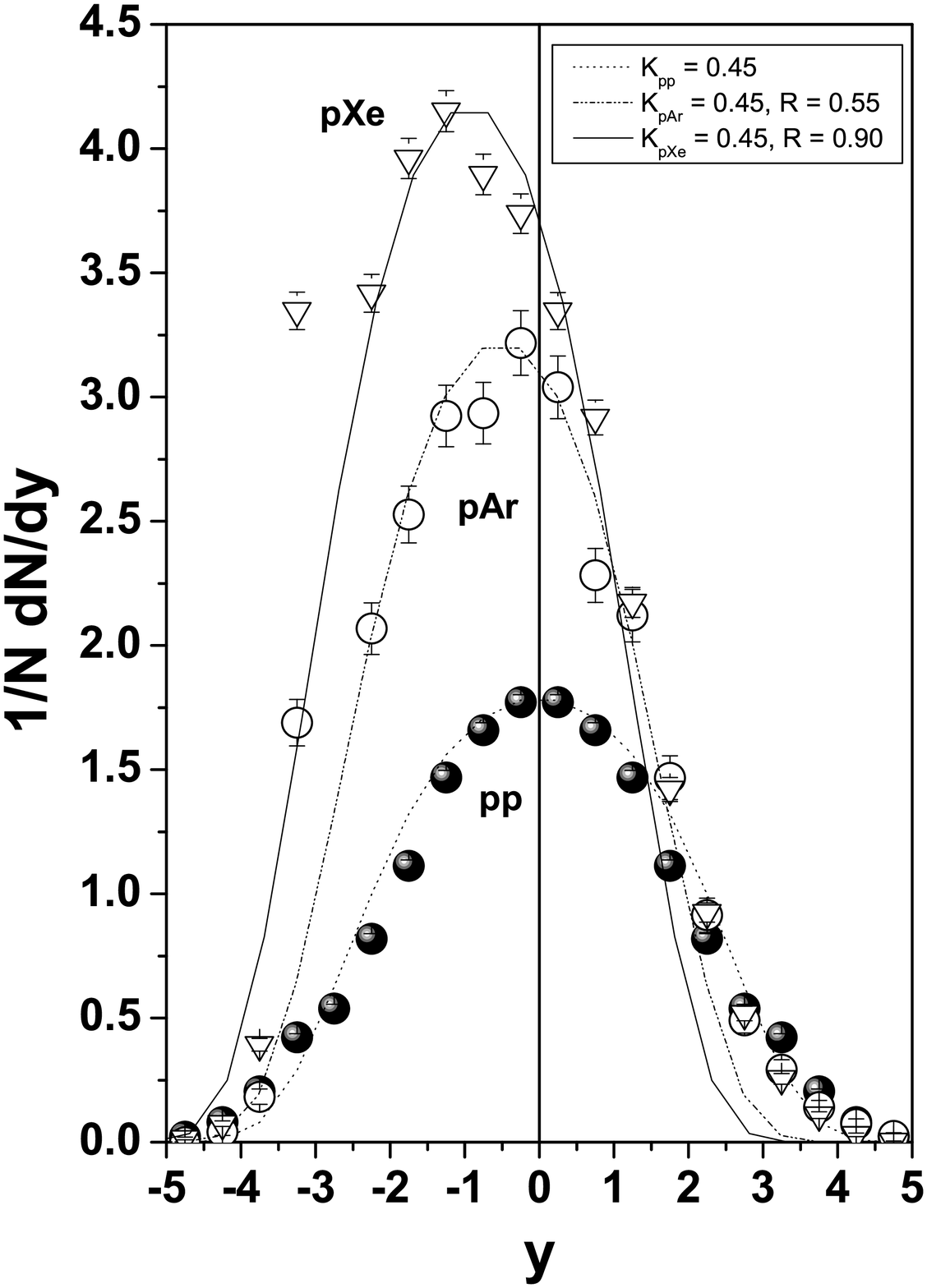, width=53mm}}
  \end{minipage}
\end{center}
\begin{center}
  \begin{minipage}[ht]{5.3cm}
    \centerline{\epsfig{file=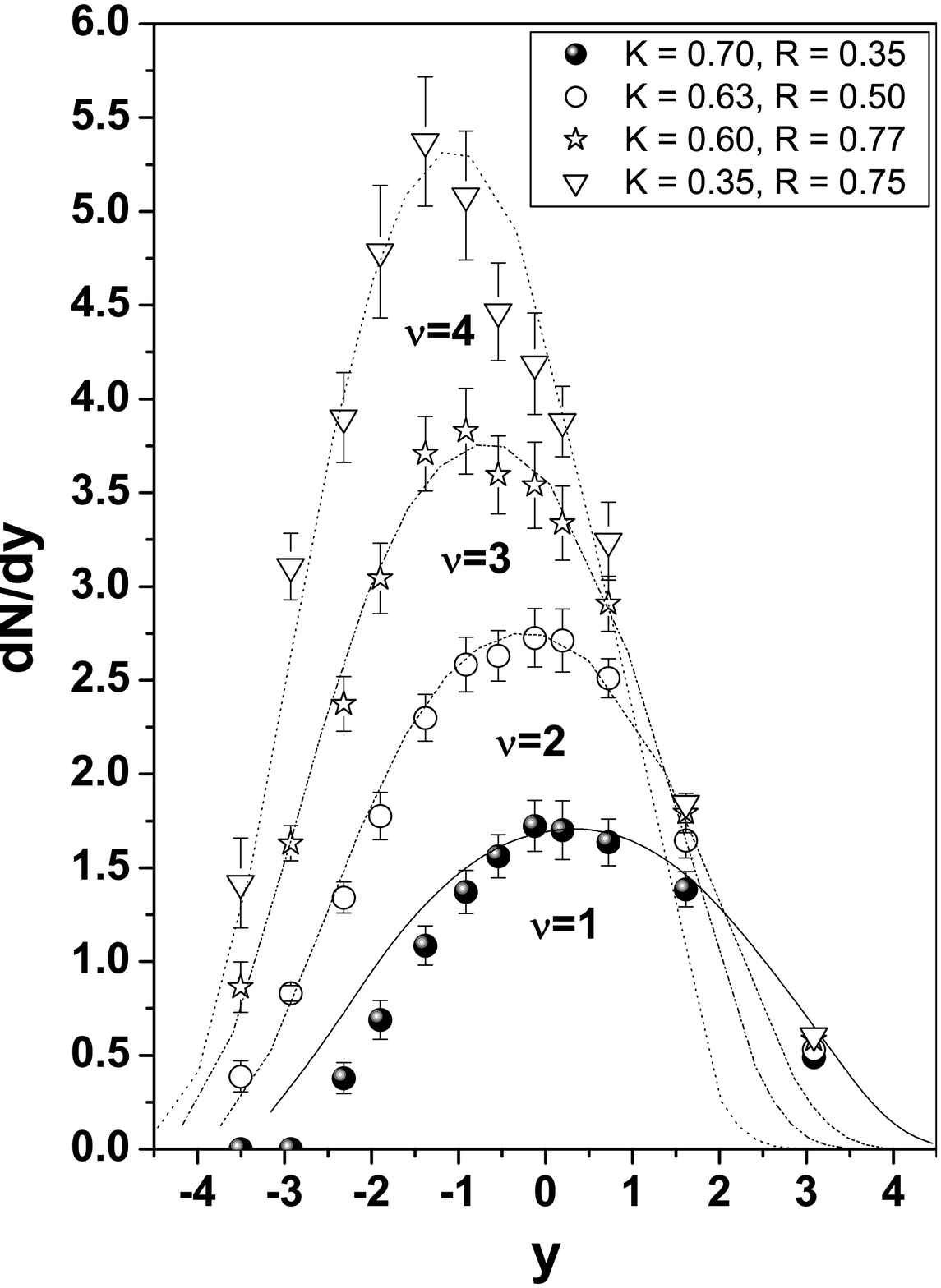, width=53mm}}
  \end{minipage}
\hfill
  \begin{minipage}[ht]{5.3cm}
    \centerline{\epsfig{file=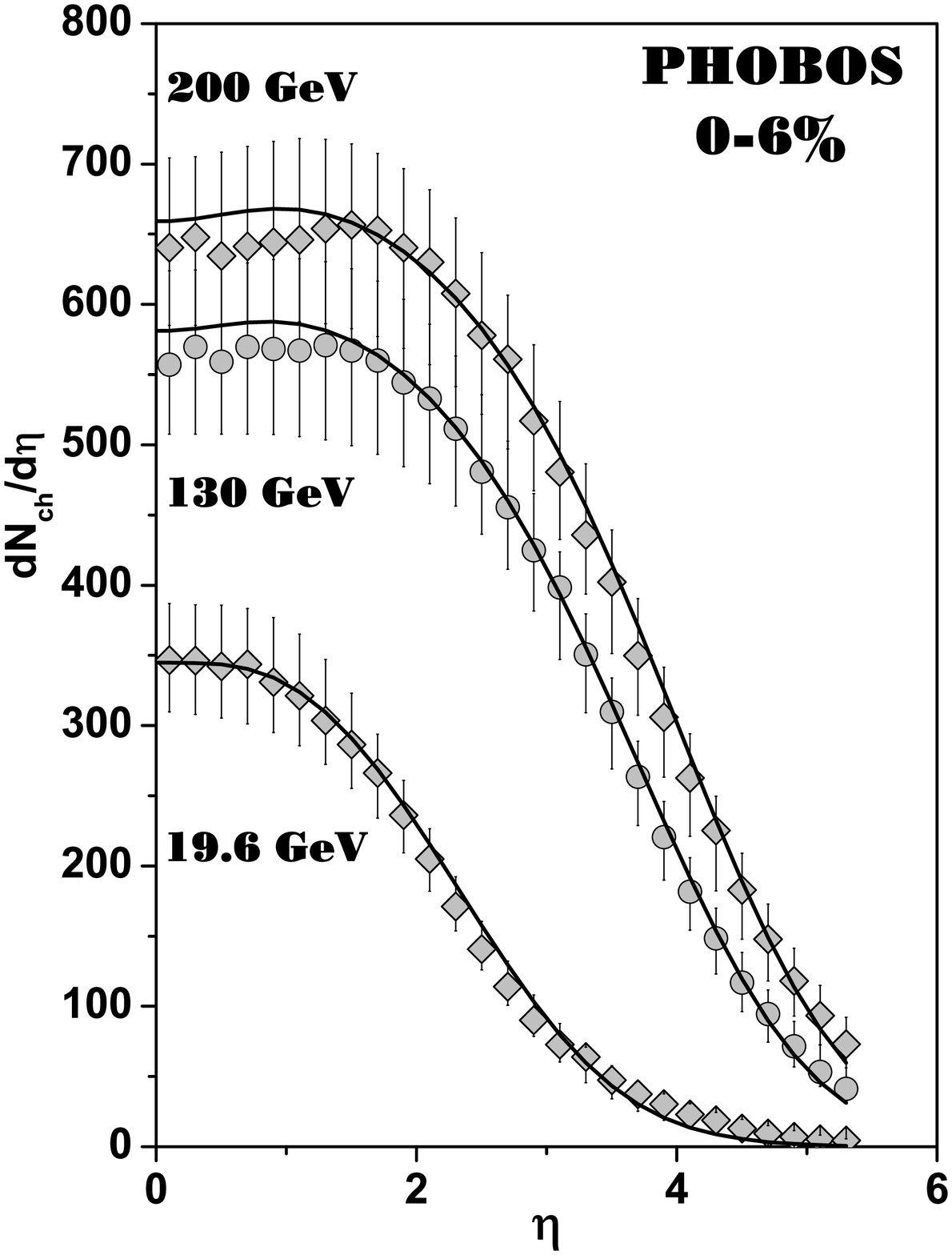, width=55mm}}
  \end{minipage}
\hfill
  \begin{minipage}[ht]{5.3cm}
    \centerline{\epsfig{file=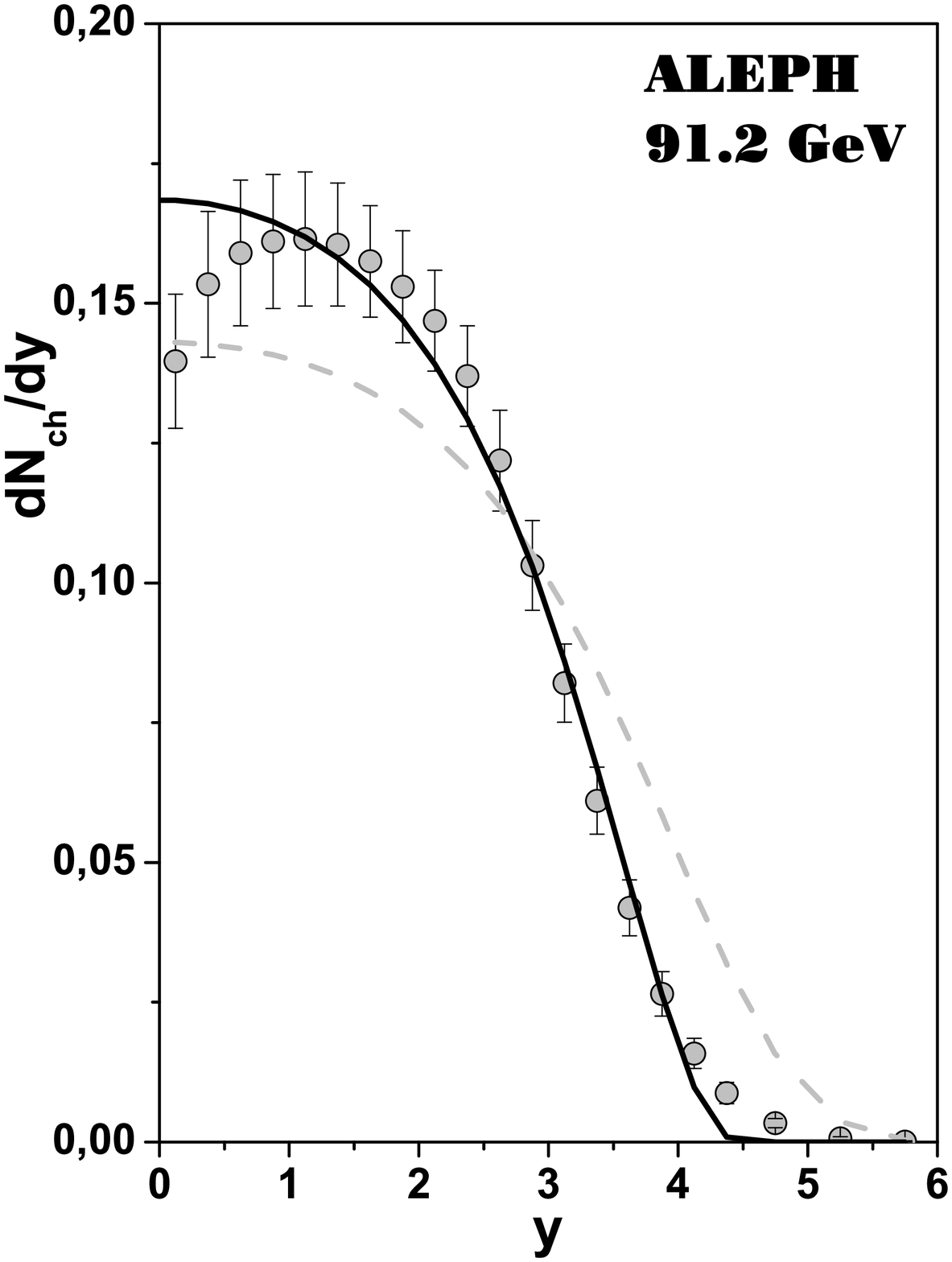, width=53mm}}
  \end{minipage}
\end{center}
\caption{Examples of application of IT approach to different kind of
data. Top-left: $dN/dy$ for $pp$ and $\bar{p}p$ collisions at different
energies, fitted parameters are inelasticity and $q=q_L$, cf.,
\protect\cite{Kq} for details. Top-middle: $p_T$ distributions at
different energies \protect\cite{MaxEntq} (fitted parameter is $q=q_T$
which remains much smaller than $q_L$. Top-right: $dN/dy$ in the so
called "tube model" of $pA$ collisions, here $K$ and $R$ are fraction of
energies of, respectively, incoming nucleon and "tube" used in
hadronization process, cf., \protect\cite{pA} for details. Bottom-left:
$p-(\nu)$nucleonic rapidity distributions for different values of hit
nucleons in the tube model \protect\cite{pA}. Notice that data on
collisions of incoming proton with nuclear nucleon ( $\nu =1$) disagree
with data for $pp$ collision at the same energy (lowest curve at
upper-left panel) and cannot be therefore fitted - the probably reason is
that "nucler nucleon" is a mixture of $p$ and $n$. Bottom-middle:
$dN/d\eta$ distributions in $AA$ collisions \protect\cite{AA}.
Bottom-right: attempt to fit $e^+e^-$ data - as one can see they cannot
be fitted properly by IT the method, cf. \protect\cite{AA}. Here the best
fit is for $q<1$ which therefore limits available phase space.}
\label{fig:examples}
\end{figure}

In Fig. \ref{fig:examples} we provide some selected examples of
application of IT to describe single particle distributions in different
reactions. In general fits are good and prefer values of the
nonextensivity parameter $q>1$ (except in the attempt to fit $e^+e^-$
data, cf. last panel of Fig. \ref{fig:examples}, where $q<1$ is preferred
and even then one cannot reproduce all features of data) and therefore
its meaning here is worth of comment. As has been shown in \cite{WWq} it
is given by fluctuations in the parameter $1/\Lambda$ in the exponential
distribution of the form $\sim \exp(-x/\Lambda)$,
namely\footnote{Although in \cite{WWq} fluctuations were assumed to be
described by gamma function this result is general and lead to
introduction of idea of {\it superstatistics}, cf., \cite{Beck}.}:
\begin{equation}
q\, =\, q_T\, =\,  1 \pm \frac{\left\langle
\left(\frac{1}{\Lambda}\right)^2\right\rangle - \left\langle
\frac{1}{\Lambda}\right\rangle^2}{\left\langle
\frac{1}{\Lambda}\right\rangle^2}. \label{eq:qfluct}
\end{equation}
When applied to $p_T$ distributions (like in the upper-right panel of
Fig. \ref{fig:examples}, cf. also \cite{Biya} and references therein) it
can be therefore connected with fluctuations of what is usually regarded
in thermodynamical descriptions of collisions as being the "temperature"
of the hadronizing system. In what concerns rapidity distributions (the
rest of Fig. \ref{fig:examples}) as was argued in \cite{Kq},
\begin{equation}
q\, =\, q_L\, = 1\, +\, \frac{1}{k}, \label{eq:qk}
\end{equation}
where $k$ is parameter characterizing (together with mean multiplicity
$\bar{n}$) the so called Negative Binomial multiplicity distribution
$P_{NB}(n)$. This is so because, as was shown in \cite{NBD},
\begin{equation}
P_{NBD}(n)\, =\, \int_0^{\infty} d\bar{n}
\frac{e^{-\bar{n}}\bar{n}^n}{n!}\cdot
         \frac{\gamma^k \bar{n}^{k-1} e^{-\gamma \bar{n}}}{\Gamma (k)} =
   \frac{\Gamma(k+n)}{\Gamma (1+n) \Gamma (k)}\cdot
   \frac{\gamma^k}{(\gamma +1)^{k+n}}, \label{eq:PNBD}
\end{equation}
where
\begin{equation}
\gamma = \frac{k}{\langle \bar{n}\rangle}\qquad {\rm and}\qquad
\frac{1}{k} = \frac{\sigma^2(\bar{n})}{\bar{n}^2} =
\frac{\sigma^2(n)}{\langle n\rangle^2}- \frac{1}{\langle n\rangle},
\label{eq:k}
\end{equation}
i.e., $P_{NBD}(n)$ arises from Poisson distribution by fluctuating its
mean multiplicity $\bar{n}$ using gamma distribution. In general $q_L$
dominates in the collision process (being of the order of $q_L\sim 1.2$
in comparison to $q_T\sim 1.02$). One also observes tendency that $q_T$
is smaller for bigger hadronization systems \cite{Kq,Biya} what agrees
with suggestion \cite{AA} that reflecting fluctuations of temperature
$q_T = 1+1/C$ where $C$ is the heat capacity of the system and as such is
expected to grow with the collision volume.

\section{Summary}

The question arises: what is the advantage of the IT method? To answer it
let us first notice that in examples presented in Fig. \ref{fig:examples}
IT was represented by most probable and least biased distribution
(\ref{eq:pi}) describing allocation of given (known) number of
secondaries in a (longitudinal) phase space defined by a given (or
assumed, by using parameter of inelasticity $K$) available energy. It
means then that \cite{MaxEnt,Kq} in (\ref{eq:pi}) one has only one term,
$k=1$, with $F_1$ being energy of secondary under consideration and the
lagrange multiplier $\lambda_1 = \beta$, being the inverse of
"temperature" (understood in the sense mentioned before). This formula is
apparently identical with formula used by statistical models of
hadronization, however, here both $\beta$ and $Z$ are not free parameters
(see \cite{Kq} for discussion and references) but are obtained from the
constraint equations (here energy conservation and
normalization)\footnote{Similar situation is for the formula
(\ref{eq:qpi}). Actually, in both cases normalization $Z$ and $Z_q$ can
be exchanged for the requirement of properly and exactly reproducing the
multiplicity of secondaries produced in a given event.}. The parameters
fitted is the energy available for hadronization (i.e., inelasticity
parameter $K$, cf., \cite{MaxEnt,Kq}, in some cases like $e^+e^-$
collisions and some $AA$ collisions they are fixed by the requirement of
experiment) and parameter $q$, which as we argue, defines the amount of
dynamical, intrinsic fluctuations present in the hadronizing system. In
case that data cannot be fitted by this method we should add some other
constraints (as in $pA$ case \cite{pA}), or turn to the true dynamical
description (as is probably the case with $e^+e^-$ collisions).

Let us end with remark that IT does not solve our dynamical problems. On
the other hand it is the only approach which allows us to select a {\it
minimal number} of {\it indispensable} hypothesis (assumptions) needed to
reproduce experimental data under consideration. In this approach any new
hypothesis are allowed only when discrepancy with some new (or
additional) experimental results occur. The choice of the form of
information entropy (here represented by parameter $q$) offers additional
flexibility because, as was stressed here, $q$ summarizes many possible
dynamical effects (out of which we have stressed here
fluctuations\footnote{It can be argued that it summarizes also effects of
correlations, especially those arising from production of resonances
\cite{UWW,Biya}. Recently question of connecting $q$ with other forms of
correlations has been discussed in \cite{qCorr}.}). Therefore assumptions
tested by using methods of IT can serve as ideal starting point to build
any dynamical model of hadronization process.\\

Two of us (OU and GW) would like to acknowledge support obtained from the
Bogolyubov-Infeld program in JINR. Partial support of the Polish State
Committee for Scientific Research (KBN) (grant
621/E-78/SPUB/CERN/P-03/DZ4/99 (GW)) is also acknowledged.



\begin{thebibliography}{99}

\bibitem{Chao} Y.-A.Chao, {\sl Nucl. Phys.} {\bf B40} (1972) 475.

\bibitem{MaxEnt} G.Wilk and Z.W\l odarczyk, {\sl Phys. Rev.} {\bf
                 D43} (1991) 794.

\bibitem{ENT} F.~Tops\oe, {\sl Physica} {\bf A340} (2004) 11.

\bibitem{T} C.~Tsallis, in {\it Nonextensive Statistical Mechanics
            and its Applications}, S.Abe and Y.Okamoto (Eds.),
            Lecture Notes in Physics LPN560, Springer (2000), in
            {\sl Physica} {\bf A340} (2004) 1 and {\sl Physica}
            {\bf A344} (2004) 718, and references therein.

\bibitem{TMT} A.M.~Teweldeberhan, H.G.~Miller and R.~Tegen, {\sl Int. J.
              Mod. Phys.} {\bf E12} (2003) 395.

\bibitem{Kq} F.S.~Navarra, O.V.~Utyuzh, G.~Wilk and Z.~W\l odarczyk,
             {\sl Phys. Rev.} {\bf D67} (2003) 114002.

\bibitem{MaxEntq} F.S.~Navarra, O.V.~Utyuzh, G.~Wilk and
                  Z.~W\l odarczyk, {\sl Physica} {\bf A340} (2004) 467.

\bibitem{pA} O.V.Utyuzh, G.Wilk and Z.W\l odarczyk, {\it Multiparticle
             production processes from the Information Theory point of view},
             hep-ph/0503048, to be published in {\it Acta Phys. Hung.}
             {\bf A} (HIP) (2005).

\bibitem{AA} F.S.~Navarra, O.V.~Utyuzh, G.~Wilk and
             Z.~W\l odarczyk, {\sl Physica} {\bf A344} (2004) 568.

\bibitem{WWq} G.Wilk and Z.W\l odarczyk, {\sl Phys. Rev. Lett.} {\bf
              84} (2000) 2770; {\sl Chaos, Solitons and Fractals}
              {\bf 13} (2002) 581 and {\sl Physica} {\bf A305} (2002)
              227.
\bibitem{Beck} C.~Beck and E.G.D.~Cohen, {\sl Physica} {\bf A322} (2003) 267.

\bibitem{Biya} M.~Biyajima, M.~Kaneyama, T.~Mizoguchi and G.~Wilk,
               {\sl Eur. Phys. J.} {\bf C40} (2005) 243.

\bibitem{NBD} P.~Carruthers and C.C.~Shih, {\sl Int. J. Mod. Phys.} {\bf A4}
              (1989) 5587.

\bibitem{UWW} O.V.~Utyuzh, G.~Wilk and Z.~W\l odarczyk, {\sl J. Phys.}
              {\bf G26} (2000) L39.

\bibitem{qCorr} C.Tsallis, M. Gell-Mann and Y.Sato, {\it Asymptotically
                scale-invariant occupancy of phase space makes the entropy Sq
                extensive}, cond-mat/0502274; T.S.Bir\'{o}, G.Purcsel,
                G.Gy\"{o}rgyi, A.Jakov\'{a}c and Z.Schramd, {\it Power-law
                tailed spectra from equilibrium}; nucl-th/0510008, talk given
                at QM2005, Budapest (and references therein).

\end{thebibliography}
\end{document}